\documentclass[10pt]{iopart}
\usepackage{graphicx}
\usepackage{color}
\usepackage{bm}
\usepackage{xcolor}
\usepackage{array}
\expandafter\let\csname equation*\endcsname\relax
\expandafter\let\csname endequation*\endcsname\relax
\usepackage{amsthm}
\usepackage{amsmath}
\usepackage{amssymb}
\usepackage{url, hyperref}

\usepackage{iopams}  
\begin{document}

\title[Ground-state properties of Dipolar Bose polarons]{Ground-state properties of Dipolar Bose polarons}

\author{L. A. Pe\~{n}a Ardila and T. Pohl}

\address{Department of Physics and Astronomy, Aarhus University, DK 8000 Aarhus C, Denmark}
\ead{luis@phys.au.dk}
\vspace{10pt}
\begin{indented}
\item[]September 2018 
\end{indented}

\begin{abstract}
We consider a quantum impurity immersed in a dipolar Bose Einstein condensate and study the properties of the emerging polaron. We calculate the energy, effective mass and quasi-particle residue of the dipolar polaron and investigate their behaviour with respect to the strength of zero-range contact and a long-range dipolar interactions among the condensate atoms and with the impurity. While quantum fluctuations in the case of pure contact interactions typically lead to an increase of the polaron energy, dipole-dipole interactions are shown to cause a sign reversal. The described signatures of dipolar interactions are shown to be observable with current experimental capabilities based on quantum gases of atoms with large magnetic dipole moments such as Erbium or Dysprosium condensates.
\end{abstract}

%
\vspace{2pc}
\maketitle
%
\ioptwocol

\section{Introduction}\label{intro}

The influence of an impurity onto its surrounding represents a paradigmatic problem in many-body physics, and, for example, plays a vital role for the transport properties of semiconductors \cite{Mahan2000book,Alexandrov2007,Devreese2009}, molecular spectra in superfluid helium \cite{Whaley1994,Vilesov98,Schmidt2017-2} or superconductivity \cite{Dagotto1994,Shen2008,Devereaux2011}. Central to the understanding of these phenomena is the concept a quasi-particle -- the polaron, which emerges from the interaction between the impurity and its environment. Since introduced by Landau and Pekar more than 80 years ago \cite{Pekar46,Pekar48}, the rich behaviour of polarons continues to play an important role in physics, both theoretically and experimentally.
The ability to control interactions in ultracold atomic quantum gases now makes it possible to investigate quantum impurity problems with unprecedented detail and accuracy. Indeed substantial theoretical~\cite{Prokofev2008,Punk2009,Massignan2012,Massignan2014,Rath2013,Li2014,Christensen2015,Levinsen2015,Wei2015,Ardila2015,Ardila2016,Grusdt15,Shchadilova2016,Grusdt2017,Bruun18,Nielsen18} and experimental~\cite{Schirotzek2009,Scazza2017,Baarsma2012,Koschorreck2012,Kohstall2012,Spethmann2012,Widera2015} efforts have been directed towards understanding the physics of polarons in cold gases composed of Fermionic  but also Bosonic  atoms. Yet, only recently experiments succeeded \cite{Jorgensen2016,Hu2016} to probe the properties of the Bose polaron for a wide range of interactions, extending from the weak-coupling into the strong-coupling regime \cite{Ardila18}. Central to all of these experimental breakthroughs is the ability to employ atomic Feshbach resonances \cite{Chin2010} for tuning the effective strength of collisional zero-range interactions.

At the same time, the exploration of finite-range interactions has been another recent frontier in cold atom research. Currently, cold gases of polar molecules \cite{ye2008,Ye2008-2,ye2017}, highly excited Rydberg atoms \cite{Gross16A,Gross16B} as well as magnetic ground state atoms \cite{Griesmaier05,Ferlaino2012,Lev2011,Lev2015_1,Lev2015_2,Kadau16,FerlainoRoton} are among the most promising systems and under active investigation. Since the first successful Bose Einstein condensation of chromium atoms in \cite{Griesmaier05} more recent experiments with erbium and dysprosium atoms make it now possible to observe the profound effects of strong dipole-dipole interactions in Bosonic quantum gases. This includes the spontaneous formation of meta-stable spatial patterns \cite{Kadau16,Wenzel2017,Ferrier2018}, the generation of quantum droplets \cite{Ferrier16,Schmitt16,Chomaz2016,Ferrier2018b} and the observation of the roton-maxon mode of the collective excitations of an axially elongated dipolar condensate \cite{FerlainoRoton}.

 Motivated by these recent experimental advances we study here the properties of the dipolar Bose polaron. In contrast to previous work \cite{Ling2014} on anisotropy effects in three-dimensional condensates, here we consider quasi two-dimensional geometries and investigate how the roton-maxon excitation spectrum as well as dipolar impurity interactions affect the ground state properties of the emerging polaron. 
 
The article is organized as follows. In section \ref{secII} we introduce the considered system and outline its theoretical treatment in terms of the Fr\"ohlich model which describes the system in the limit of weak interactions. Its perturbative solution and the resulting polaron ground state energy are determined in section \ref{secIII}, followed by a discussion of the associated quas-particle residue and the effective mass of the polaron. Implications of our calculations for potential experiments and future perspectives are discussed in sections \ref{secIV} and \ref{secV}.

\begin{figure}
\begin{center}
\includegraphics[width=8.5cm]{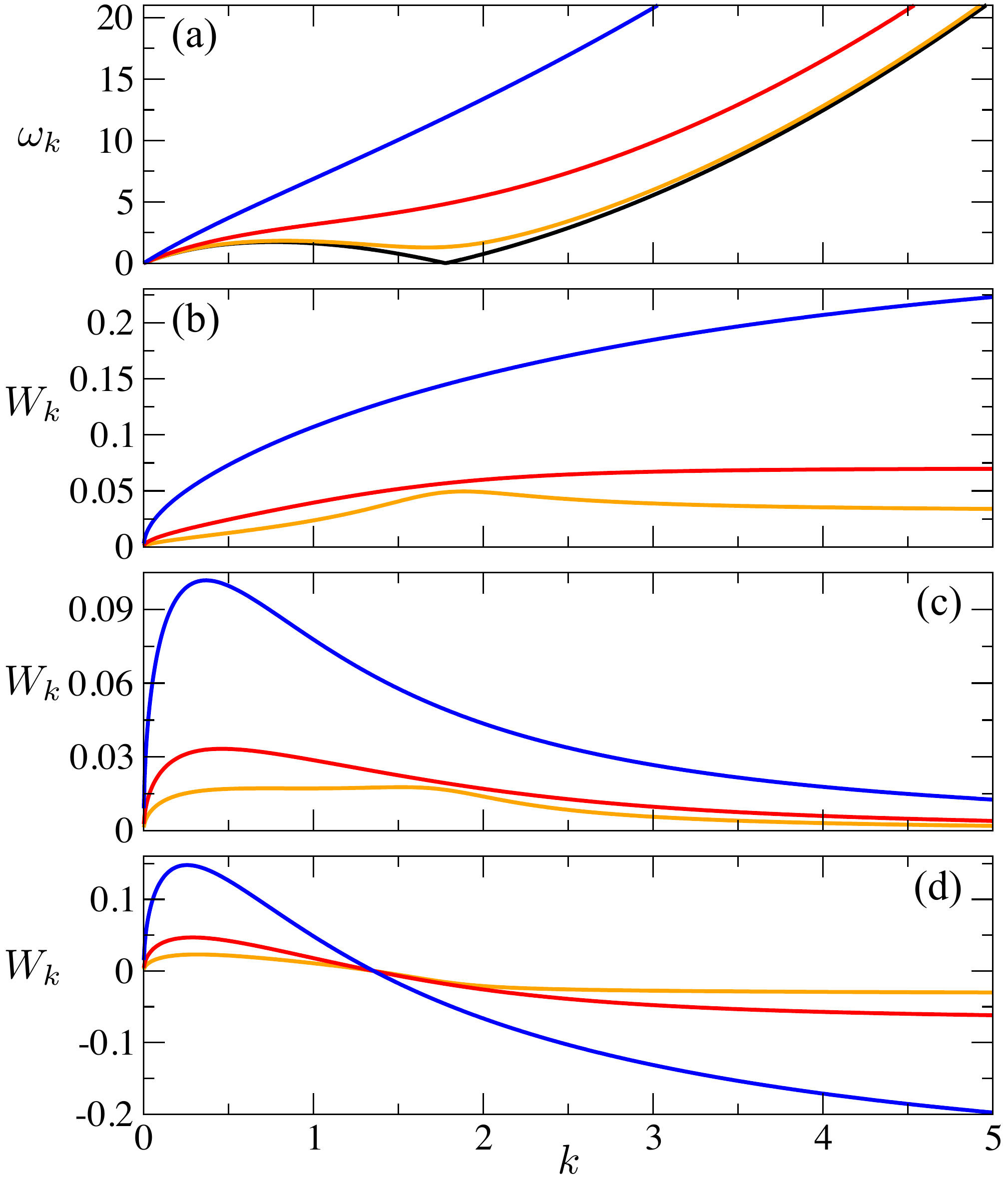}
\caption{(color online). (a) Dispersion relation, $\omega_{k}$, of the quasi two-dimensional dipolar Bose Einstein condensate with $\nu=1$, a dimensionless density of $n=125$, a dimensionless dipolar length of $a_B^{(d)}=0.007$ and different zero-range scattering lengths of $a_B^{(c)}=4a_B^{(d)}$ (blue line), $a_B^{(c)}=a_B^{(d)}$ (red line), $a_B^{(c)}=0.45a_B^{(d)}$ (orange line), and $a_B^{(c)}=0.405a_B^{(d)}$ (black line). Roton softening occurs in the latter case. Panels (b)-(d) show the effective impurity-condensate interaction, $W_k$ [cf. eqs.(\ref{eq:Frohlich2}) and (\ref{eq:W})], for these parameters using the same color coding. The impurity scattering length is $a_I^{(c)}=a_B^{(c)}$, while the different panels show results for different impurity dipolar lengths $a_I^{(d)}=0$ (b), $a_I^{(d)}=a_I^{(c)}$ (c), and $a_I^{(d)}=2a_I^{(c)}$ (d). The figure shows dimensionless quantities, where energies and lengths are scaled by $\hbar^2/(m_B d^2)$ and $d/\sqrt{2}$, respectively.}
\label{fig1}
\end{center}
\end{figure}

\section{The system and its effective Hamiltonian}\label{secII}
We consider an impurity of mass $m_{I}$ immersed in a dipolar Bose-Einstein condensate of particles with mass $m_{B}$. The interaction $V_B$ between the condensate atoms and the impurity-boson interaction $V_I$ can both be written as
\begin{equation} \label{eq:potential}
V_{\mu}(\mathbf{R})=g_{\mu}^{(c)}\delta(\mathbf{R})+\frac{g_{\mu}^{(d)}}{4\pi\left|\mathbf{R}\right|^{3}}\left[1-3\frac{z^{2}}{\left|\mathbf{R}\right|^{2}}\right],
\end{equation}
where the index $\mu={ I},{ B}$ refers to either the impurity ($I$) or the condensate atoms ($B$).
The first term in equation~(\ref{eq:potential}) accounts for zero-range collisions that can be described by an effective contact interaction potential with a strength $g_{B}^{(c)}=4\pi \hbar^{2}a_{B}^{(c)}/m_{B}$ and $g_{I}^{(c)}=2\pi\hbar^{2}a_{I}^{(c)}/m_{r}$, respectively, 
and where $\delta(\mathbf{R})$ denotes the Dirac delta function of the interatomic distance vector ${\bf R}$. Moreover,  $a_{B}^{(c)}$  and  $a_{I}^{(c)}$  are the respective \textit{s-}wave scattering lengths and $m_{r}=m_{B}m_{I}/(m_{B}+m_{I})$ is the reduced mass.  The second term in equation~(\ref{eq:potential}) describes the dipole-dipole interaction with respective dipolar coupling strengths  $g_{B}^{(d)}=12\pi \hbar^{2}a_{B}^{(d)}/m_{B}$ and $g_{I}^{(d)}=6\pi \hbar^{2}a_{I}^{(d)}/m_{r}$, expressed in terms of the  boson-boson  and impurity-boson dipolar lengths  $a_{B}^{(d)}$ and $a_{I}^{(d)}$, respectively  ~\cite{Lahaye09}.

Since the partially attractive nature of the dipole-dipole interaction can cause condensate collapse~\cite{Santos00}, we focus here on a stable two-dimensional geometry with strong particle confinement in one direction ($z$-axis) and polarized dipoles along the same axis. For strong harmonic confinement one can consider a static Gaussian density profile $\rho(z)=(\pi d^{2})^{-1/2}\exp\left(-\frac{z^{2}}{d^{2}}\right)$, along the $z$-axis in order to derive an effective two-dimensional Hamiltonian~\cite{Fischer06} that describes the system dynamics in the two-dimensional plane $\mathbf{r}=(x,y)$. Upon scaling energies by $\frac{\hbar^{2}}{m_{B}d^{2}}$ and lengths by $d/\sqrt{2}$, the effective Hamiltonian of the system can be written as
\begin{eqnarray}  
\hat{H}=&-&\nu \nabla_{\mathbf{r}_I}^{2}-{\normalcolor \int d\mathbf{r}{\color{red}{\color{green}{\normalcolor \hat{\Phi}^{\dagger}(\mathbf{r})\nabla_{\mathbf{r}}^{2}\hat{\Phi}(\mathbf{r})}}}}\nonumber\\
&+&\frac{1}{2}\int d\mathbf{r}d\mathbf{r}'{\normalcolor {\color{green}{\normalcolor \hat{\Phi}^{\dagger}(\mathbf{r})\hat{\Phi}^{\dagger}(\mathbf{r'})U_{B}(\mathbf{r'}-\mathbf{r})\hat{\Phi}(\mathbf{r'})\hat{\Phi}(\mathbf{r})}}}\nonumber\\
&+&\int d\mathbf{r}\hat{\Phi}^{\dagger}(\mathbf{r})U_{I}\left(\left|\mathbf{r}-\mathbf{r}_{I}\right|\right)\hat{\Phi}(\mathbf{r}),
\label{eq:HH}
\end{eqnarray}
where $\nu=m_{B}/m_{I}$ is the mass ratio for the condensate atoms and the impurity, and $\mathbf{r}_I$ denotes the position of the impurity. Considering only a single impurity, we treat the impurity in first quantization, while the condensate is described in second quantisation in terms of the bosonic creation operators $\hat{\Phi}^\dagger(\mathbf{r})$ for the condensate atoms. The effective interaction potentials $U_B({\bf r})$ and $U_I({\bf r})$ are readily obtained from equation~(\ref{eq:potential}) by integrating over the assumed transverse Gaussian density profile \cite{Fischer06}. The result is particularly simple in momentum space representation 
\begin{equation}
U_\mu(k)=2\sqrt{\pi}(1+m_\mu/m_B)\left[\bar{a}_{\mu}^{(c)}+3\bar{a}^{(d)}_\mu f (k)\right],
\label{eq:FourierBB}
\end{equation}
where $ f(k)=\left[\frac{2}{3}-\sqrt{\pi}{k}e^{{k}^{2}}\mathtt{erfc}({k})\right]$, $\bar{a}_{\mu}^{(c)}$ and $\bar{a}^{(d)}_\mu$ denote the dimensionless scattering lengths and dipolar lengths in units of $d/\sqrt{2}$. 

\begin{figure}[t!]
\begin{center}
\includegraphics[width=8.5cm]{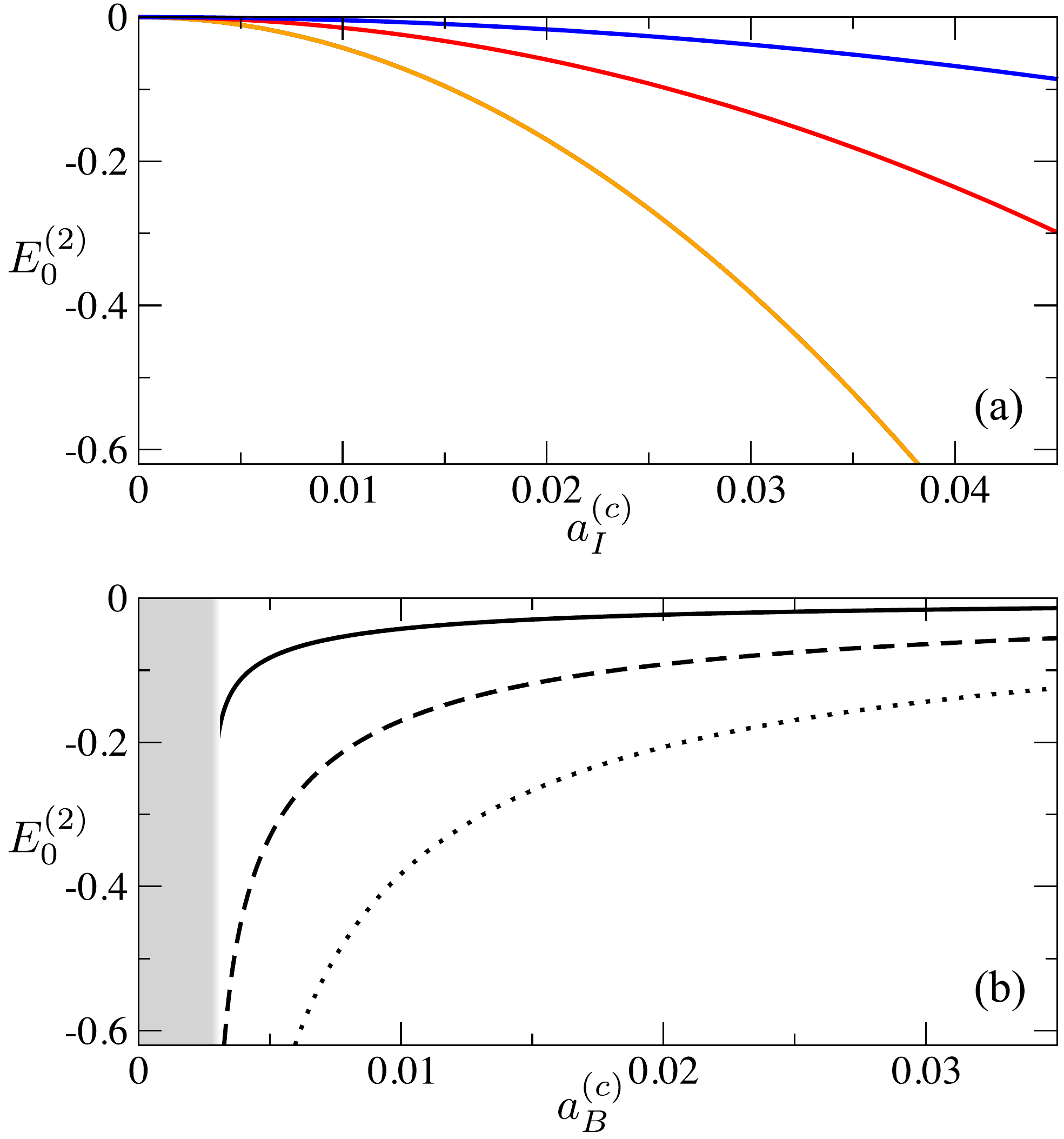}
\caption{(a) Second order contribution to the polaron energy, $E^{(2)}_0$, describing the leading order effects of quantum fluctuations, as a function of the dimensionless  impurity scattering length $a_I^{(c)}$ for a condensate density with $\nu=1$, $n=125$, a dipolar length $a_B^{(d)}=0.007$ of the condensate atoms and a dipolar length $a_I^{(d)}=a_I^{(c)}$ of the impurity. The different lines show results for different values of $a_B^{(c)}=4a_B^{(d)}$ (blue line), $a_B^{(c)}=a_B^{(d)}$ (red line), and $a_B^{(c)}=0.45a_B^{(d)}$ (orange line). Panel (b) shows $E^{(2)}_0$ as a function of $a_B^{(c)}$ for $a_I^{(c)}=0.02$ (solid line), $a_I^{(c)}=0.04$ (dashed line), and $a_I^{(c)}=0.06$ (dotted line). The figure shows dimensionless quantities, where energies and lengths are scaled by $\hbar^2/(m_B d^2)$ and $d/\sqrt{2}$, respectively. The gray region represents the value for which the bosonic dispersion of the dipolar Bose gas has imaginary phononic modes which drives the system into collapse.}
\label{fig2}
\end{center}
\end{figure}

We proceed by applying Bogoliubov theory to describe the Bose Einstein condensate in terms of its macroscopically occupied ground state, with the two-dimensional particle density $n$, and Bogoliubov excitations created by the associated bosonic operators $\hat{\alpha}_{\mathbf{k}}^{\dagger}$. In the limit of weak interactions this brings equation~(\ref{eq:HH}) into the Fr\"{o}hlich-Bogoliubov form \cite{Tempere2009}, $\hat{H}\approx \hat{H}_{0}+\hat{H}_{\rm int}$, where 
\begin{equation}
\hat{H}_{0}=-\nu\nabla_{\mathbf{r}}^{2}+\sum_{\mathbf{k}}\omega_{\mathbf{k}}\hat{\alpha}_{\mathbf{k}}^{\dagger}\hat{\alpha}_{\mathbf{k}}.
\label{eq:Frohlich1}
\end{equation}
is determined by the dispersion $\omega_{k}=k\sqrt{k^{2}+2n{U}_B({k})}$ of the Bogoliubov excitations. The resulting interaction Hamiltonian  
\begin{eqnarray} 
\hat{H}_{int}&=&E_{\rm mf}+\frac{\sqrt{{n}}}{{L}}\sum_{\mathbf{k}}W_{\mathbf{k}}e^{-i\mathbf{k\cdot r_{I}}}\left(\hat{\alpha}_{\mathbf{k}}+\hat{\alpha}_{\mathbf{-k}}^{\dagger}\right).
\label{eq:Frohlich2}
\end{eqnarray}
contains a constant mean field energy  $E_{\rm mf}=2\sqrt{\pi}(1+\nu)(a_{I}^{(c)}+2a_{I}^{(d)}){n}$ and the effective impurity-boson interaction 
\begin{equation}
W_{{k}}=(1+\nu)\sqrt{8\pi}\sqrt{\frac{k^{2}}{\omega_{k}}}\left(a_{I}^{(c)}+3a_{I}^{(d)}f(k)\right).
\label{eq:W}
\end{equation}
While the presence of the roton minimum in the dispersion spectrum tends to enhance the effects of quantum fluctuations, the Bogoliubov approximation remains valid as long as the value of the roton minimum exceeds $\sqrt{128\pi^{2}n^{2}\left(a_{B}^{(c)}\right)^{3}}$ \cite{Shlyapnikov2013}. Moreover, the interacting Hamiltonian~(\ref{eq:Frohlich2}), corresponding to the Fr\"ohlich model, neglects higher order terms in the mode operators $\hat{\alpha}_{\mathbf{k}}$, which provides a good approximation in the weak interaction limit \cite{Christensen2015} considered in this work.

The characteristic shape of the effective interaction potential is shown in figure~\ref{fig1}. In the absence of dipole-dipole interactions $W_{k}$ monotonously approaches a constant, $(1+\nu)\sqrt{8\pi} a_{I}^{(c)}$, while the modified dispersion relation of dipolar condensates can induce a maximum or minimum of $W_{\mathbf{k}}$ but does not affect its asymptotic value. A dipolar impurity, however, can render the effective interaction sign-indefinite and lead to negative asymptotic interactions, $W(k\rightarrow\infty)=(1+\nu)\sqrt{8\pi}\left(a_{I}^{(c)}-a_{I}^{(d)}\right)$. Note that such behaviour is also responsible for the emergence of roton excitations in dipolar condensates, which where recently observed in Er condensates \cite{FerlainoRoton}. In the present case, the properties of dipolar Bose polarons are determined by both the emergence of roton-type condensate excitations as well as the dipolar impurity interactions.

\section{Calculation of the polaron energy}\label{secIII}
Since the Fr\"ohlich model for the Bose polaron constitutes a weak-coupling theory it suffices to perform a perturbative expansion to leading order in the impurity interaction strengths. The total state of the system, $|{\bf P}, 0\rangle=|{\bf P}\rangle|0\rangle$, in the absence of interactions describes the impurity with momentum ${\bf P}$ and the condensate groundstate $|0\rangle$ containing no excitations. The interaction with the impurity couples the ground state to an excited state $1_{\bf k}$ with a single occupied Bogoliubov mode and a total energy $E_{\bf k}^{(0)}=\nu(\mathbf{P-k})^2+\omega_{\mathbf{k}}$, accounting for momentum conservation.  Therefore the first order energy correction, $E^{(1)}=\langle{\bf P},0|\hat{H}_{int}|{\bf P},0\rangle=E_{\rm mf}$ simply corresponds to the mean field energy shift defined above, while the second order contribution follows from 
\begin{eqnarray}\label{eq:E2P}
E^{(2)}(\mathbf{P})&=&\sum_{{\bf k}\neq {\bf 0}}\frac{|\langle {\bf P}-{\bf k},1_{\bf k}|\hat{H}_{int}|{\bf P},0\rangle |^{2}}{E_{n}^{(0)}-E_{k}^{(0)}}\nonumber\\
&=&\frac{n}{(2\pi)^2}\int \frac{|W_{k}|^{2}d{\bf k}}{\nu P^{2}-\nu\left(\mathbf{P-k}\right)^{2}-\omega_{\mathbf{k}}}\;.
\end{eqnarray}
Upon expanding this expression to second order in ${\bf P}$, one can read off the shift 
\begin{equation}
E_0^{(2)}=-\frac{n}{2\pi}\int_0^\infty dk\frac{k\left|W_{k}\right|^{2}}{\nu k^{2}+\omega_{k}}.
\label{eq:E}
\end{equation}
of the polaron energy along with the correction to the effective impurity mass due to the formation of its surrounding screening cloud.

As in the case of pure contact interactions, the energy correction displays an ultraviolet divergence, that is usually dealt with through a renormalization procedure \cite{Tempere12} or by introducing an upper momentum cutoff \cite{Grusdt15}. In the present situation however, the integral becomes convergent for strong dipolar interactions $a_{I}^{(d)}=a_{I}^{(c)}$. In this case the short range part of the dipolar interaction precisely cancels the collisional interaction such that $W_{k \rightarrow \infty} = 0$, thus ensuring the convergence of the integral in equation~(\ref{eq:E}).

\begin{figure}[t!]
\begin{center}
\includegraphics[width=8.5cm]{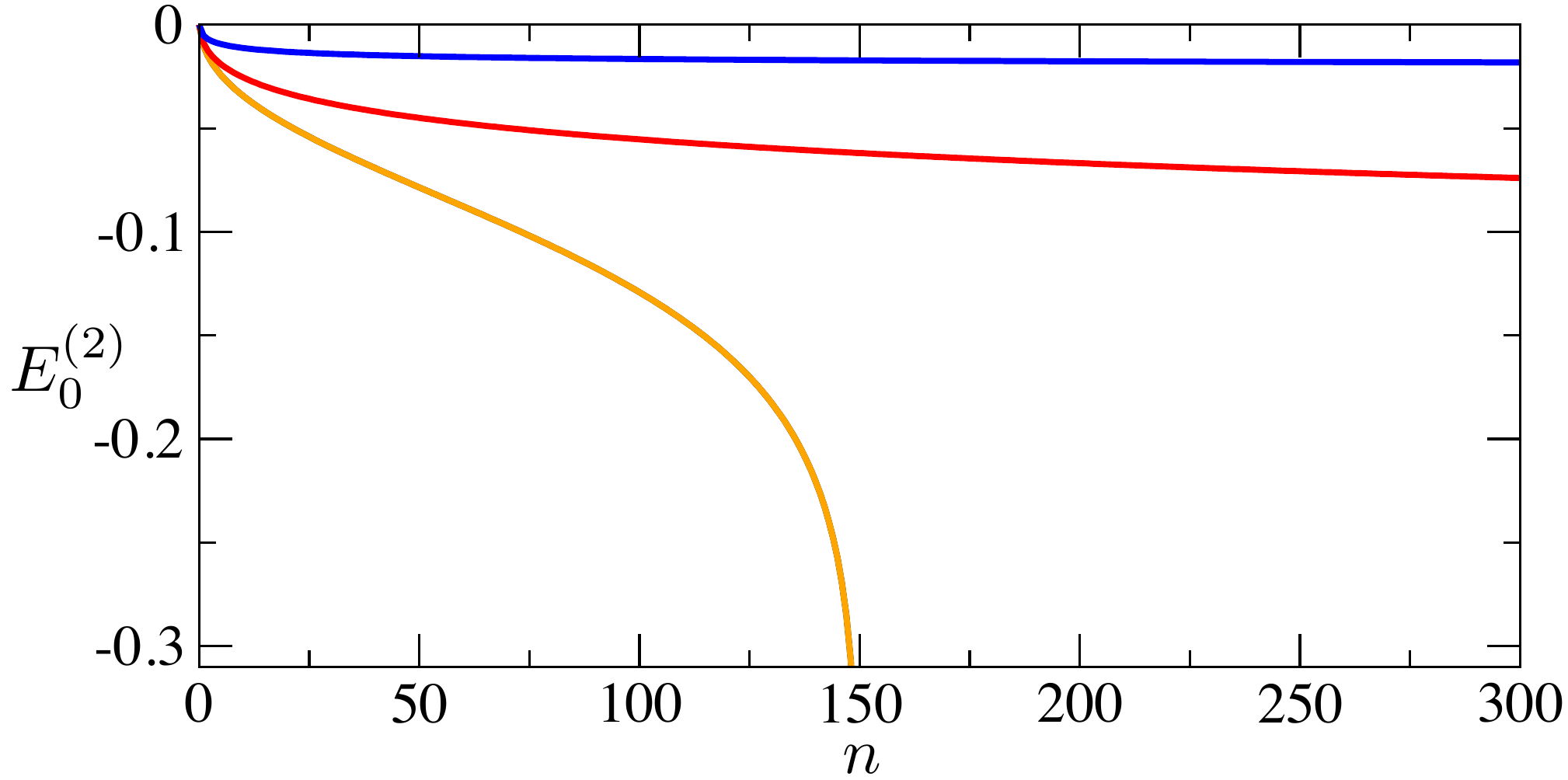}
\caption{Second order contribution to the polaron energy, $E^{(2)}_0$, as a function of the dimensionless density $n$ for $\nu=1$, a dipolar length $a_B^{(d)}=0.007$ of the condensate atoms, and a dipolar length $a_I^{(d)}=a_I^{(c)}=0.02$ of the impurity. The different lines show results for different values of $a_B^{(c)}=4a_B^{(d)}$ (blue line), $a_B^{(c)}=a_B^{(d)}$ (red line), and $a_B^{(c)}=0.45a_B^{(d)}$ (orange line). The figure shows dimensionless quantities, where energies and lengths are scaled by $\hbar^2/(m_B d^2)$ and $d/\sqrt{2}$, respectively.}
\label{fig3}
\end{center}
\end{figure}

Figure~\ref{fig2}(a) shows the polaron energy $E_0^{(2)}$ as a function of the scattering length of the condensate atoms for different values of $a_{I}^{(c)}$. In contrast to the case of pure contact interactions, quantum fluctuations yield a negative contribution, $E_0^{(2)}<0$, to the total energy. Moreover, we see that their effect gets strongly enhanced close to the roton softening. This enhancement results from the generation of roton excitations with  momenta around $k_{\rm rot}$ [cf. figure~\ref{fig1}(a)] whose energy cost decreases rapidly close to the roton instability of the condensate. This is seen more directly when considering the energy as a function of $a_B^{(c)}$ as shown in figure~\ref{fig2}(b). Indeed, the energy shift increases rapidly close to the critical value for the roton instability. A similar behaviour can be found with respect to the condensate density as shown in figure~\ref{fig3}, where the energy shift grows rapidly and eventually diverges around the critical density for  roton softening.
As a result, one can effectively enter the strong impurity-coupling regime, despite having weak impurity interactions.

\section{Effective polaron mass and quasi-particle residue}\label{secIV}
\begin{figure}
\includegraphics[width=8.5cm]{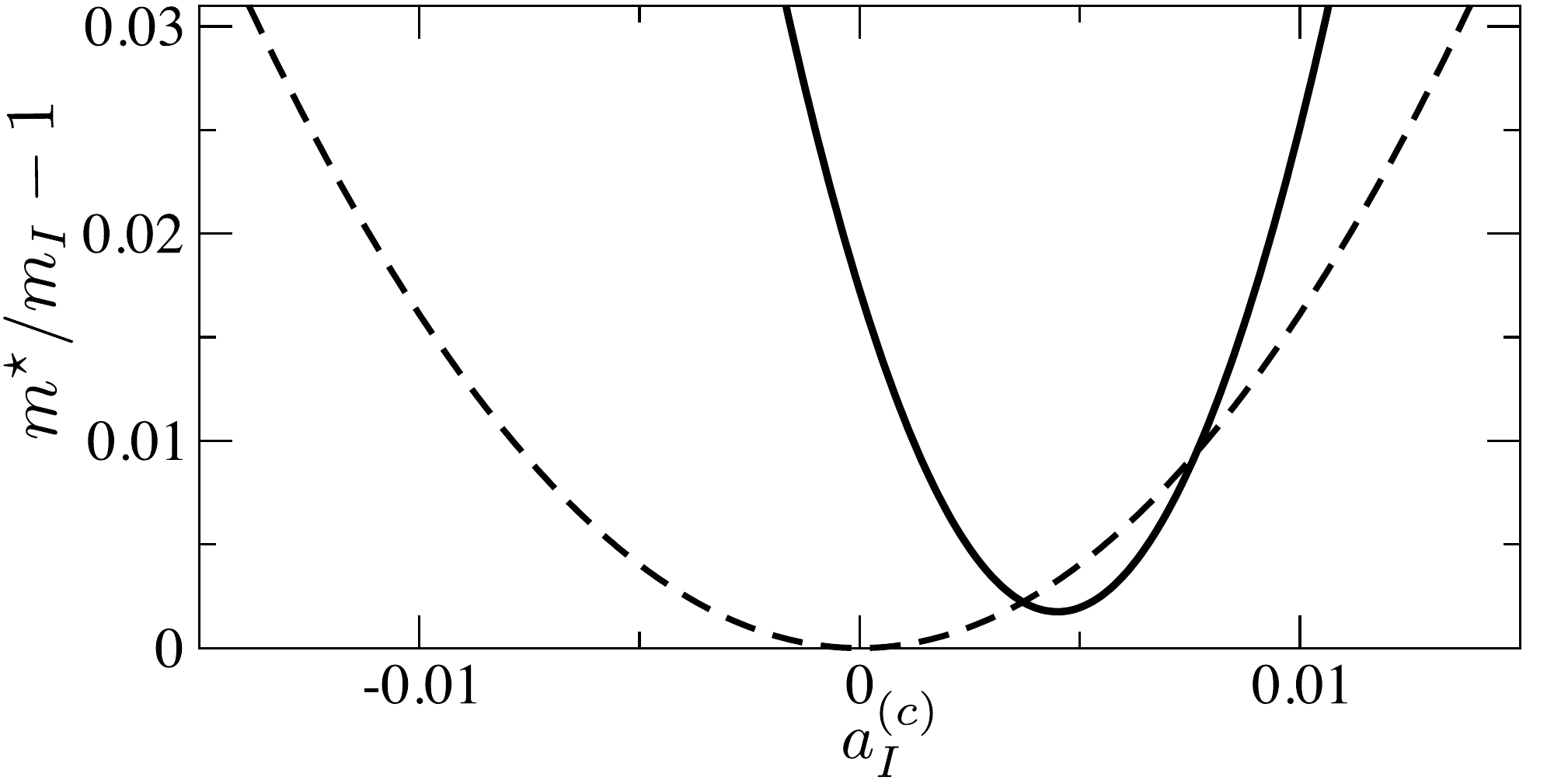}
\caption{The effective mass of the polaron as a function of $a_{I}^{(c)}$ for $\nu=1$, a density $n=125$  and a scattering length $a_B^{(c)}=0.0035$ of the condensate atoms. The different lines are for dipolar lengths $a_B^{(d)}=a_I^{(d)}=0$ (dashed line) and $a_B^{(d)}=a_I^{(d)}=0.007$ (solid line). The figure shows dimensionless quantities, where energies and lengths are scaled by $\hbar^2/(m_B d^2)$ and $d/\sqrt{2}$, respectively.}
\label{fig4}
\end{figure}

Two additional important quantities characterize the major properties of the Bose polaron, namely its effective mass $m^{\star}$ and its quasiparticle residue $Z$.  As discussed above, the effective mass, $m^\star$ of the dipolar polaron can be obtained directly from expanding equation~(\ref{eq:E2P}) for small impurity momenta and is given by
\begin{equation}
\frac{m^{\star}}{m_I}=1+\frac{2n}{\pi}\int dkk^{3}\frac{|W_{k}|^{2}}{\left(\nu k^{2}+\omega_{k}\right)^{3}}.
\label{eq:M}
\end{equation}
This equation describes the increase of the effective  mass of the polaron relative to the bare impurity mass due to the formation of a screening cloud around the impurity which inevitably reduces its mobility. 
As a result the effective mass monotonously increases with the magnitude of the scattering length in the  case of pure contact interactions for both positive and negative values of $a_I^{(c)}$ \cite{Christensen2015}.
Closely related to this behavior is the quasiparticle residue $Z$ which describes the fraction of the bare impurity state that is contained in the state of the polaron. A stronger screening or dressing by the surrounding condensate may therefore be expected to cause a decreasing residue just as it leads to an increasing effective mass. The quasi-particle residue is readily obtained from the polaron wavefunction and is given by 
\begin{equation}
\sqrt{Z}=1-\frac{n}{4\pi}\int dk \frac{k|W_{k}|^{2}}{\left(\nu k{}^{2}+\omega_{k}\right)^{2}}
\label{eq:Z}
\end{equation}
 to leading order in the impurity interaction.
The parabolic dependence of the effective polaron mass on the scattering length $a_I^{(c)}$ in the absence of dipolar interactions is shown in figure~\ref{fig4}. Dipole interactions, on the other hand, shift the minimum to finite values of $a_I^{(c)}$. Being also determined by the screening of the impurity, the quasi-particle residue shows qualitatively similar behaviour, i.e. a shift of its maximum to finite values of the scattering length $a_I^{(c)}$. 

\begin{figure}[t!]
\begin{center}
\includegraphics[width=8.5cm]{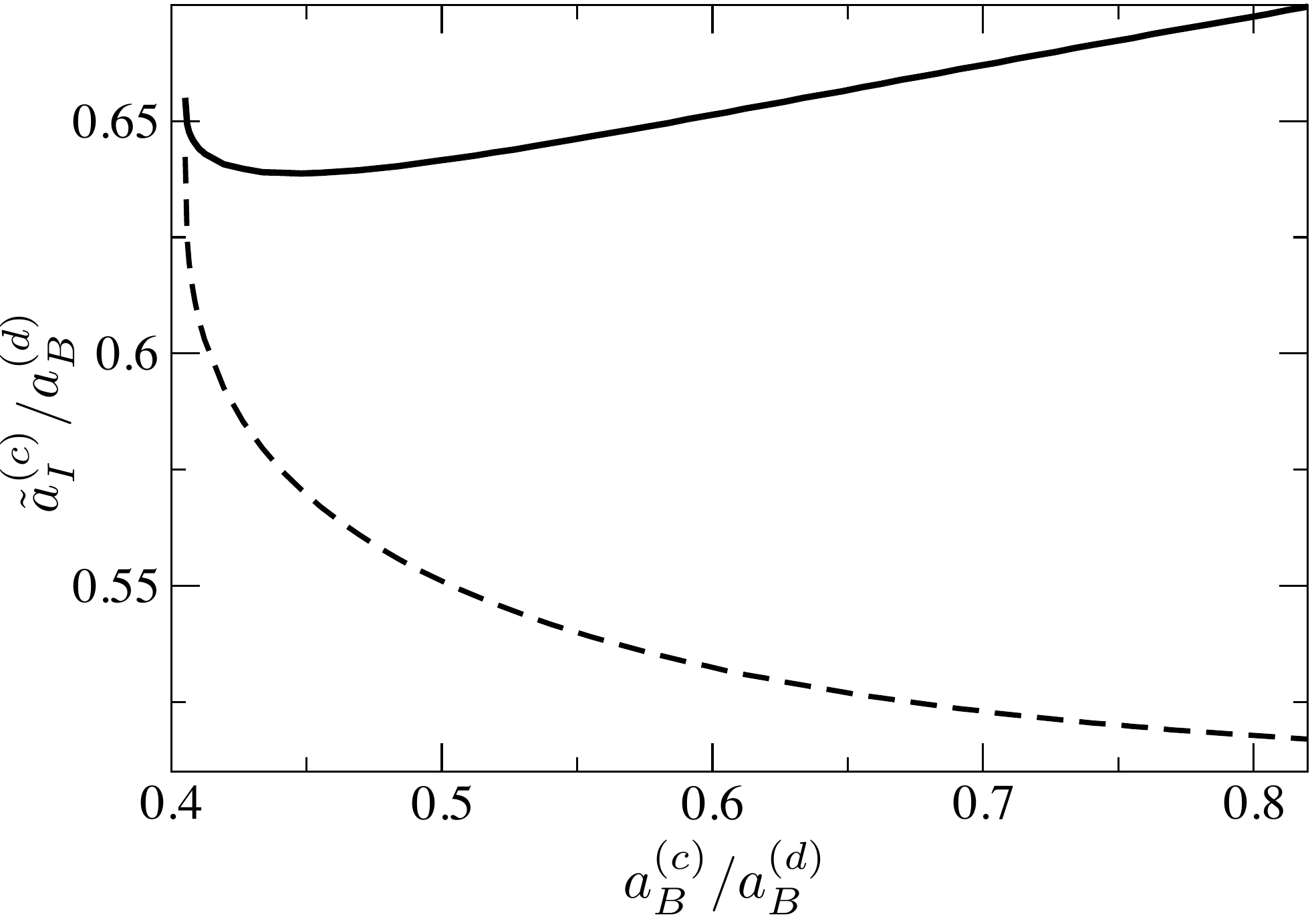}
\caption{Position $\tilde{a}_I^{(c)}$ of the minimum of the effective mass (solid line) as indicated in figure~\ref{fig4} as a function of $a_B^{(c)}$ and otherwise identical parameters. The dashed line shows the position of the minimum with respect to the quasi-particle residue $Z$. The figure shows dimensionless quantities, where energies and lengths are scaled by $\hbar^2/(m_B d^2)$ and $d/\sqrt{2}$, respectively.}
\label{fig5}
\end{center}
\end{figure}

\begin{figure}[b!]
\begin{center}
\includegraphics[width=8.5cm]{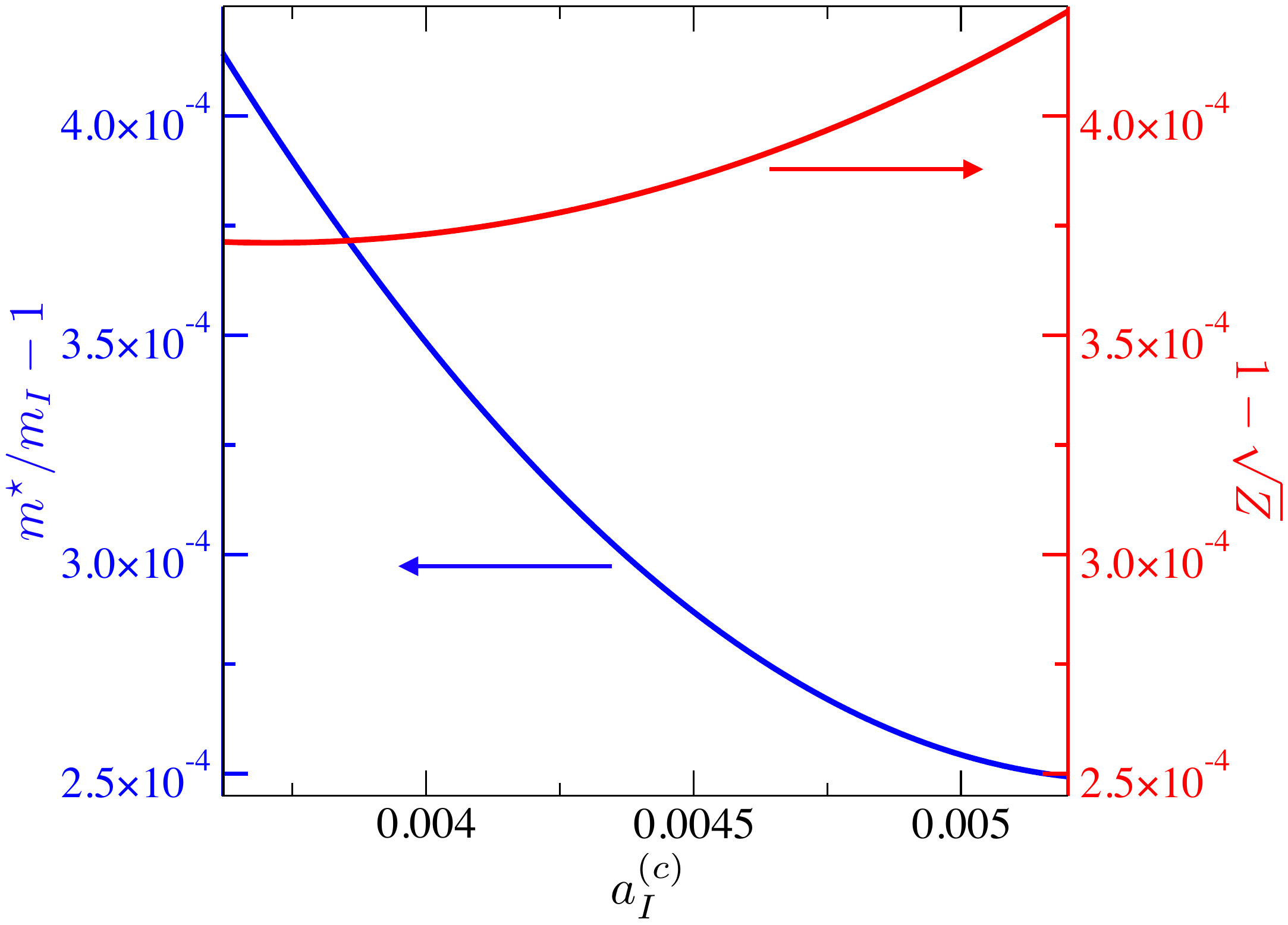}
\caption{Effective mass (blue line) and quasi-particle residue (red line) of the polaron as a function of the impurity scattering length $a_I^{(c)}$. The remaining parameters are $\nu=1$, $n=125$, $a_B^{(d)}=a_I^{(d)}=0.007$, and $a_B^{(c)}=2a_B^{(d)}$. The figure shows dimensionless quantities, where energies and lengths are scaled by $\hbar^2/(m_B d^2)$ and $d/\sqrt{2}$, respectively.}
\label{fig6}
\end{center}
\end{figure}

However, the position, $\tilde{a}_I^{(c)}$ of the extrema of the quasi-particle residue and the effective mass differ appreciably, as shown in figure~\ref{fig5}. While the extrema in both cases result from a partial cancellation of the competing effects of the zero-range interaction and the finite range dipole-dipole interactions  between the impurity and the condensate atoms, $m^\star$ and $\sqrt{Z}$ are affected differently by this compensation, leading to the different values of $\tilde{a}_I^{(c)}$ for the effective mass and quasi-particle residue, respectively. Interestingly, this can lead to the unusual situation where stronger dressing of the impurity due to an increasing scattering length $a_I^{(c)}$ causes its quasi-particle residue to drop, while at the same time leading to a decreasing effective mass of the polaron, as shown in figure~\ref{fig6}. In order to ensure the applicability of the Fr\"ohlich model and the validity of the perturbative treatment we have chosen parameters for which $m^{\star}/m-1$ and  $Z-1$ remain small. Yet, the found characteristic effects of dipolar interactions are expected to persist qualitatively and lead to a further enhancement of the effective mass for stronger impurity interactions. 

 \section{Relation to experiments}\label{secV}
We finally discuss prospects for observing the described signatures of dipole-dipole interactions in future polaron experiments. The most promising candidates to date are Bose Einstein condensates of Erbium and Dysprosium atoms which feature large magnetic dipole moments and have been studied in a number of recent experiments \cite{Kadau16,Wenzel2017,Ferrier2018,Ferrier16,Schmitt16,Chomaz2016,Ferrier2018b,FerlainoRoton}. Here, the generation and subsequent probing of impurities would be possible by optical or microwave coupling to another hyperfine state as recently demonstrated in potassium Bose Einstein condensates ~\cite{Jorgensen2016}. Alternatively, one could employ different atomic species or different isotopes as employed in recent polaron experiments with mixtures of potassium and rubidium atoms \cite{Hu2016}. In fact, Bose-Einstein condensation of $^{160}$Dy \cite{Lev2015_1}, $^{162}$Dy \cite{Lev2015_2} and  $^{164}$Dy \cite{Ferrier16} has been achieved in a number of recent experiments. Dy condensates feature a dipolar interaction length of $132a_0$, where $a_0$ is the Bohr radius. For a typical thickness of $d/\sqrt{2}=1 \mu$m of the quasi two-dimensional condensate \cite{Dalibar2006} this yields a dimensionless scattering length of  $a_{B}^{(d)}=0.007$, as used in the calculations above. Taking a typical peak density of $5\times10^{13}$cm$^{-3}$ these parameters give the employed dimensionless two-dimensional density of $n=125$. Moreover, the considered scattering lengths $a_{B}^{(c)}$ and $a_{I}^{(c)}$ are also well within the range of experimentally accessible values and can be accurately tuned via magnetic Feshbach resonances \cite{Lev2014,Lev2011,Julienne2015}.

\section{Conclusions}\label{secVI}
In conclusion, we have investigated the properties of dipolar impurities immersed in a dipolar Bose Einstein condensate. Starting from the Fr\"ohlich Hamiltonian for the two-dimensional condensate we have determined the energy of the emerging polaron as well as its effective mass and quasiparticle residue to second order in the interaction strength. We showed that dipolar interactions qualitatively affect the properties of the polaron, which for example can give rise to negative energy corrections due to quantum fluctuations. While we have focussed here on impurity interactions with $a_{I}^{(c)}=a_{I}^{(d)}$ to ensure asymptotically vanishing impurity interactions, $W(k)$, the case of $a_I^{(c)}\approx-2a_{I}^{(d)}$ appears to be particularly interesting as it allows to suppress the mean field energy $E_{\rm mf}\sim a_{I}^{(c)}+2a_{I}^{(d)}$. Similar to recent work on Bosonic mixtures with pure contact interactions \cite{Petrov2015,Cabrera2018} or dipolar  condensates \cite{Ferrier16,Schmitt16,Chomaz2016}, the competition between the thereby reduced mean-field energy and $E^{(2)}(\mathbf{0})$ may potential induce the formation of quantum droplets but seeded by the impurity, which for multiple impurities could cause structured or even ordered states to emerge well outside the roton instability of the underlying BEC. The observation of this effect as well as the behavior predicted in the present work likely requires strong interactions, calling for a strong-coupling for the dipolar Bose-polaron. It may lead also to interesting effects such as self-localisation of the polaron. In view of the peculiar effects of the roton dispersion, touched upon in this work, the nature of the emerging effective interaction \cite{Casteels2013,Naidon2016,Dehkharghani2017,Camacho2017,Bruun18} between dipolar Bose polarons suggests itself as an interesting question for future work.

\section*{Acknowledgments}
We thank to Georg Bruun, Russell Bisset and Wim Casteels for valuable discussions. This work was supported by the DNRF through a Niels Bohr Professorship.

\section*{References}
\bibliography{DiBoPo.bib}
\end{document}